\documentclass[preprint,12pt]{elsarticle}

\usepackage{amssymb}
\usepackage{amsmath}

\usepackage{lineno}

\begin{document}

\begin{frontmatter}



\title{Multiple quantum NMR in solids as a method of determination of Wigner--Yanase skew information}


\author[icp]{S.~I.~Doronin}
\author[icp]{E.~B.~Fel'dman}
\author[icp,msu]{I.~D.~Lazarev}
\address[icp]{Institute of Problems of Chemical Physics of Russian Academy of Sciences, \\ Chernogolovka, Moscow Region, Russia 142432}
\address[msu]{Faculty of Fundamental Physical-Chemical Engineering, Lomonosov Moscow State University, GSP-1, Moscow, Russia 119991}

\begin{abstract}
A connection of the Wigner--Yanase skew information and multiple quantum (MQ) NMR coherences is considered at different temperatures and evolution times of nuclear spins with dipole-dipole interactions in MQ NMR experiments in solids.
It is shown that the Wigner--Yanase skew information at temperature $T$ is equal to the double second moment of the MQ NMR spectrum at the double temperature for any evolution times.
A comparison of the many--spin entanglement obtained with the Wigner--Yanase information and the Fisher information is conducted.
\end{abstract}




\end{frontmatter}



\section{Introduction}
\label{sec:1}
The Wigner--Yanase skew information~\cite{1,2,3,4} together with the Fisher information~\cite{5,6} allow the development of powerful methods for the investigation of entanglement, including many--particle entanglement~\cite{7,8}.
Further investigations of many--particle entanglement require the development of corresponding experimental methods.
In particular, it was shown~\cite{7,9} that a lower bound on the quantum Fisher information~\cite{5,6} coincides with the double second moment of the spectrum of multiple quantum (MQ) coherences.
As a result, the lower bound on the quantum Fisher information can be found in MQ~NMR experiments~\cite{10},
in cold--atom experiments, including experiments with Bose--Einstein condensates, ultracold atoms in cavities, and trapped ions~\cite{11,12,13,14,15}.
Using the properties of the quantum Fisher information one can obtain the number of the entangled particles (spins) in the system under consideration~\cite{7}
and even find the dependence of the number of the entangled spins on the temperature~\cite{9}.

The Wigner--Yanase skew information~\cite{1,2,3,4} is also connected with the spectrum of MQ coherences.
In particular, we demonstrate in the present article that the Wigner--Yanase skew information in a spin system $(s = 1/2)$ with the dipole--dipole interactions (DDI) in the MQ NMR experiment~\cite{10} at the system temperature $T$ equals the double second moment of the MQ NMR spectrum obtained at the temperature $2T$.
Using the properties of the Wigner--Yanase skew information one can investigate many--spin entanglement on the basis of the MQ NMR spectroscopy~\cite{10}.

The main aim of the present article is the development of a method of extracting the Wigner--Yanase skew information from the MQ NMR spectra.
We also compare on the simple models~\cite{8,16} the many--spin entanglement obtained both with the Wigner--Yanase information and the Fisher information.

The article is organized as follows.
In Sec.~\ref{sec:2} a short introduction to the MQ NMR spectroscopy is given.
The connection of the Wigner--Yanase skew information with the second moment of the MQ NMR spectrum is obtained in Sec.~\ref{sec:3}.
A comparison of many--spin entanglement obtained with the Wigner--Yanase and the Fisher information on simple models~\cite{8,16} is conducted in Sec.~\ref{sec:4}.
We briefly discuss our results in concluding Sec.~\ref{sec:5}.

\section{MQ NMR for solving problems of quantum informatics}
\label{sec:2}

\begin{figure}
	\includegraphics[width=0.95\linewidth]{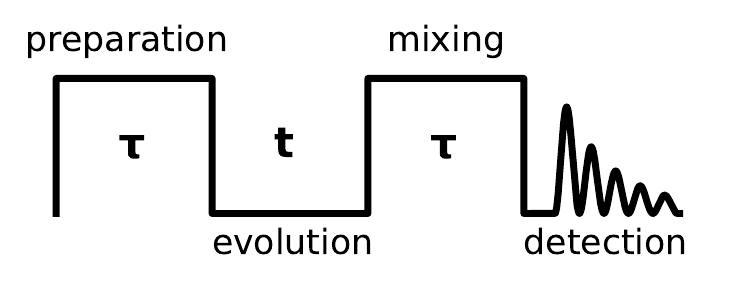}
	\caption{The basic scheme of the multiple quantum NMR experiment.}
	\label{fig:1}
\end{figure}

MQ NMR methods are widely used for solving problems of quantum informatics~\cite{17,18}.
The MQ NMR experiment consists of four distinct periods of time as depicted in Fig.~\ref{fig:1}:
preparation $(\tau)$, evolution $(t_1)$, mixing $(\tau)$, and detection $(t_2)$~\cite{10}.
MQ NMR coherences are created by a periodic multipulse sequence, consisting of
$\pm$x-pulses, irradiating the system of the preparation period~\cite{10}.
If the inverse period of the multipulse sequence significantly exceeds the local dipolar field (in frequency units)~\cite{19} then
MQ NMR dynamics can be described by the averaged nonsecular two--spin/two--quantum Hamiltonian~$H_{MQ}$~\cite{20}
\begin{equation} \label{eq:1}
        H_{MQ} = H^{(+2)} + H^{(-2)} , \quad
        H^{(\pm 2)} = -\frac{1}{2} \sum_{j<k} D_{jk} I_{j} ^\pm I_k^\pm,
\end{equation}
where $D_{jk}$ is the coupling constant between spins $j$ and $k$,
and $I_{j}^+, I_k ^-$ are the raising and lowering operators of spin $j$.
On the mixing period, the spin system is irradiated by the multiple pulse sequence with $\pm y$--pulses.
As a result, the averaged nonsecular two-spin/two quantum Hamiltonian on the mixing period equals $(-H_{MQ})$~\cite{10}.

In order to investigate the MQ NMR dynamics of the system on the preparation period~\cite{10} one should find the density matrix $\rho(t)$ by solving the Liouville evolution equation~\cite{19}

\begin{equation}
    \label{eq:2}
        i\frac{d\rho(t)}{dt} = [H_{MQ}, \rho(t)]
\end{equation}
with the initial thermodynamic equilibrium density matrix
\begin{equation}
    \label{eq:3}
        \rho(0) = \rho_\mathrm{eq} = \frac{\exp(\frac{\hbar \omega_0}{kT}I_z)}{Z},
\end{equation}
where $Z=Tr \left\{exp\left(\frac{\hbar \omega_0}{kT}I_z\right) \right\}$ is the partition function,
$\hbar$ and $k$ are the Plank and Boltzmann constants, respectively,
$\omega_0$ is the Larmor frequency,
$T$ is the temperature,
and $I_z$ is the operator of the projection of the total spin angular momentum on the $z$-axis,
which is directed along the strong external magnetic field.

Following the preparation, evolution, and mixing periods of the MQ NMR experiments and taking into account the phase increment $\phi$ of the radio-frequency pulses~\cite{10}, the resulting signal $G(\tau,\phi)$ stored as population information is
\begin{equation} \label{eq:4}
	\begin{split}
		G(\tau,\phi)
		& = Tr \left\{
			e^{iH_{MQ}\tau} e^{i\phi I_z} e^{-iH_{MQ}\tau} \rho_\mathrm{eq}
			e^{iH_{MQ}\tau} e^{-i\phi I_z} e^{-iH_{MQ}\tau}\rho_\mathrm{eq}
		\right\}
		\\
		& = Tr \left\{
			e^{i\phi I_z} \rho_\mathrm{pre}(\tau,\beta)
      e^{-i\phi I_z}\rho_\mathrm{pre}(\tau,\beta)
		\right\},
	\end{split}
\end{equation}
where
\begin{equation} \label{eq:5}
	\rho_\mathrm{pre}(\tau,\beta) = e^{-iH_{MQ}\tau}\rho_\mathrm{eq}e^{iH_{MQ}\tau}
\end{equation}
is the density matrix at the end of the preparation period.
The density matrix can be obtained from~Eqs.~(\ref{eq:2},\ref{eq:3}) and $\beta = \frac{\hbar \omega_0}{kT}$.

It is convenient to expand the density matrix $\rho_\mathrm{pre}(\tau, \beta)$ in series as~\cite{21}
\begin{equation}
    \label{eq:6}
        \rho_\mathrm{pre}(\tau,\beta) = \sum_n \rho_{\mathrm{pre},n}(\tau,\beta),
\end{equation}
where $\rho_{\mathrm{pre}, n}(\tau,\beta)$ is the contribution to the density matrix $\rho_\mathrm{pre}(\tau,\beta)$ from the MQ coherence of the n--th order.
Then the resulting signal $G(\tau,\phi)$ of the MQ NMR~\cite{10} can be rewritten as
\begin{equation} \label{eq:7}
    G(\tau, \phi) = \sum_n e^{in\phi}
        Tr\left\{\rho_{\mathrm{pre},n}(\tau,\beta)
        \rho_{\mathrm{pre},-n}(\tau,\beta) \right\},
\end{equation}
where we took into account that
\begin{equation}
    \label{eq:8}
        [I_z, \rho_{\mathrm{pre},n}] = n \rho_{\mathrm{pre},n}
\end{equation}
The normalized intensities of the MQ NMR coherences can be determined as follows
\begin{equation}
    \label{eq:9}
        J_n(\tau,\beta)= \frac{Tr\left\{\rho_{\mathrm{pre},n}(\tau,\beta)
            \rho_{\mathrm{pre},-n}(\tau,\beta)\right\}}
                {Tr(\rho^2_\mathrm{eq})}
\end{equation}
As was shown in~\cite{8},
\begin{equation}
    \label{eq:10}
        Tr(\rho_\mathrm{eq}^2) = \frac{2^N ch^N (\beta)}{Z^2},
\end{equation}
where $N$ is the number of the spins.
It was also shown that
\begin{equation}
    \label{eq:11}
        \sum_n J_n(\tau,\beta) = 1
\end{equation}
The second moment (dispersion) $M_2(\tau,\beta)$ of the distribution of the MQ NMR coherences $J_n (\tau,\beta)$ can be calculated from Eq.~(\ref{eq:7}) according to~\cite{22}
\begin{equation}
    \label{eq:12}
        M_2(\tau,\beta) = -\frac{1}{G(\tau, \beta)}
            \frac{d^2 G(\tau + t,\beta)}{dt^2}\bigg|
        _{t=0}
\end{equation}
Using Eqs.~(\ref{eq:7},\ref{eq:8},\ref{eq:12}) one can obtain
\begin{equation}
    \label{eq:13}
        M_2 (\tau,\beta) = \sum_n n^2 J_n(\tau,\beta)
\end{equation}
A lower bound on the quantum Fisher information coincides with the double second moment of Eq.~(\ref{eq:13})~\cite{7,9}.
As a result, the analysis of the temperature dependence of the second moment $M_2(\tau,\beta)$ of the distribution of the intensities of the MQ NMR coherences allows us to obtain the number of the entangled spins at different temperatures~\cite{8}.
In the following~Sec.~\ref{sec:3} we demonstrate that the Wigner--Yananse skew information is also connected with the second moment $M_2(\tau,\beta)$
and can be useful for the investigation of many-spin entanglement.

\section{The Wigner--Yanase skew information and MQ NMR}
\label{sec:3}
The Wigner--Yanase skew information is defined as~\cite{1,2,3,4}
\begin{equation}
    \label{eq:14}
        I_{WY}(\rho(\tau,\beta),I_z) = -\frac{1}{2}
            Tr([\sqrt{\rho(\tau,\beta)},\sigma_z])^2 =
                -2Tr([\sqrt{\rho(\tau,\beta)},I_z])^2,
\end{equation}
where the Pauli operator $\sigma_z=2I_z$.
Introducing the evolution operator
\begin{equation}
    \label{eq:15}
        V(\tau) = e^{iH_{MQ}\tau}
\end{equation}
and using Eq.~(\ref{eq:3}) one can write the density matrix $\rho(\tau,\beta)$ as follows:
\begin{equation}
    \label{eq:16}
        \rho(\tau,\beta) = V^+(\tau) \frac{e^{\beta I_z}}{Z}V(\tau)
\end{equation}
Now we use the evident relationship:
\begin{equation}
    \label{eq:17}
        \sqrt{\rho(\tau,\beta)} =
            \sqrt{V^+(\tau)\frac{e^{\beta I_z}}{Z}V(\tau)} =
                V^+(\tau) \frac{e^{\frac{\beta}{2}I_z}}{\sqrt{Z}}V(\tau).
\end{equation}
It can be proved by simple calculation:
\begin{equation}
    \label{eq:18}
        \sqrt{\rho}\sqrt{\rho} =
						V^+(\tau)\frac{e^{\frac{\beta}{2}I_z}}{\sqrt{Z}}
                V(\tau)V^+(\tau)\frac{e^{\frac{\beta}{2}I_z}}{\sqrt{Z}}V(\tau) =
						V^+(\tau)\frac{e^{\beta I_z}}{Z}V(\tau) =
        \rho(\tau,\beta)
\end{equation}
Then we have
\begin{equation} \label{eq:19}
    \left[I_z,\sqrt{\rho(\tau,\beta)}\right]
    = \left[I_z, \sum_k \rho_k \left(\tau, \frac{\beta}{2}\right)\right]
    = \sum_k k\rho_k \left(\tau, \frac{\beta}{2}\right),
\end{equation}
and
\begin{equation} \label{eq:20}
	Tr\left[I_z,\sqrt{\rho(\tau,\beta)} \right]^2
	= Tr\left\{\sum_{k,k'}kk'
		\rho_k\left(\tau,\frac{\beta}{2}\right)
		\rho_{k'}\left(\tau,\frac{\beta}{2}\right)
	\right\}
	= \sum_k k^2 J_k\left(\tau,\frac{\beta}{2}\right).
\end{equation}
Finally, one can obtain that
\begin{equation} \label{eq:21}
    I_{WY}\left(\rho(\tau, \beta), I_z\right)
    = 2\sum_k k^2 J_k\left(\tau, \frac{\beta}{2}\right)
    = 2M_2\left(\tau, \frac{\beta}{2}\right)
\end{equation}
Thus, we obtain an important observation.
If the spin system is investigated with MQ NMR at the temperature $T\sim\beta^{-1}$ then the Wigner--Yanase skew information equals to the double second moment of the distribution of the intensities of the MQ NMR coherences at the temperature $2T \sim 2\beta^{-1}$ at any time during the spin evolution.

The Wigner-Yanase skew information is connected with the second moment of the distribution of the MQ NMR coherences analogously to the Fisher information.
We compare these informations in the following Section~\ref{sec:4}.

\section{The comparison of the many-spin entanglement obtained with the Wigner--Yanase information and the Fisher information}
\label{sec:4}

\begin{figure}
	\includegraphics[width=0.95\linewidth]{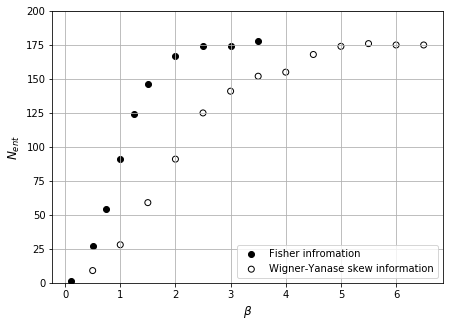}
	\caption{
		The dependence of the number of the entangled spins on the inverse temperature $\beta = \frac{\pi \omega_0}{kT}$;
		black circles - the results are obtained with the Fisher information;
		open circles - the results are obtained with the Wigner--Yanase information.
	}
	\label{fig:2}
\end{figure}

\begin{figure}
	\includegraphics[width=0.95\linewidth]{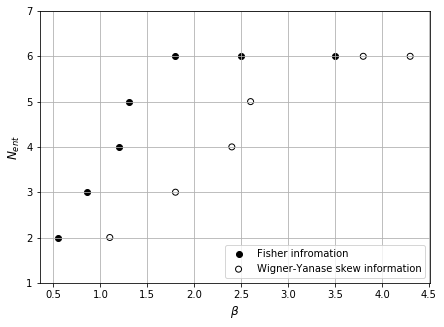}
	\caption{
		The dependence of the number $N_\mathrm{ent}$ of the entangled spins on the parameter $\beta$ (the inverse temperature) for zigzag chains consisting of six spins.
	}
	\label{fig:3}
\end{figure}

The Wigner--Yanase skew information $I_{WY}(\rho(\tau,\beta),I_z)$ and the Fisher information $I_F(\rho(\tau,\beta),I_z)$ can be used for the investigation of the many-spin entanglement.
Indeed, it is known~\cite{5,6} that if
$I_{WY}\left( \rho(\tau, \beta), I_z \right)$
or
$I_{F}\left( \rho(\tau, \beta), I_z \right)$
exceeds $mk^2 + (N - mk)^2$,
where $k, m$ are integer and $m$ is the integer part of $N/k$,
then we have $N_\mathrm{ent} = (k + 1) $~--~particle entangled spins in the system.
The informations are connected by the following restriction~\cite{3}
\begin{equation} \label{eq:22}
    I_{WY}\left(\rho(\tau,\beta), I_z\right)
    \leq I_F\left(\rho(\tau,\beta), I_z\right)
    \leq 2I_{WY}\left(\rho(\tau,\beta), I_z\right).
\end{equation}
The restrictions (22) allow us to hope that the obtained results for the number of the entangled spins are not very different.
For the comparison we used the model~\cite{23} of a nonspherical nanopore filled with a gas of spin-carrying atoms (for example, xenon) or molecules in a strong external magnetic field.
This model allows the investigation of the many-spin entanglement in the spin system consisting of hundreds of nuclear spins~\cite{8}.

We investigated many-spin entanglement in the spin system, consisting of 201 spins, in a nanopore both with
the Wigner--Yanase information $I_{WY}\left(\rho(\tau, \beta), I_z\right)$
and the Fisher information $I_F\left(\rho(\tau,\beta),I_z\right)$.
In Fig.~\ref{fig:2} the dependence of the number of the entangled spins on the inverse temperature is presented.
Fig.~\ref{fig:2} demonstrates that the number of the entangled spins increases when the temperature decreases both for the Wigner--Yanase information and the Fisher information.

An analogous investigation was conducted on the model of the proton zigzag chain in a single crystal of hambergite~\cite{16,24}.
In Fig.~\ref{fig:3},  similar results on many--spin entanglement are presented for the system consisting of six spins at different temperatures for both used informations.

\section{Conclusion}
\label{sec:5}

We studied the connection of the Wigner--Yanase skew information with the second moment of the distribution of the intensities of MQ coherences in the MQ NMR experiment.
It was shown that the Wigner--Yanase skew information at the temperature $T \sim \beta^{-1}$ equals the double second momentum of the MQ NMR spectrum at the temperature $2T \sim 2\beta^{-1}$.
We compare also the results on the many--spin entanglement obtained with the Wigner--Yanase skew information and the Fisher information.

\section{Acknowledgement}
\label{sec:6}
We acknowledge funding from the Ministry of Science and Higher Education of the Russian Federation (Grant No.~075-15-2020-779).








\end{document}